# Diagnosing Unknown Attacks in Smart Homes Using Abductive Reasoning

Kushal Ramkumar*, Wanling Cai†, John McCarthy‡, Gavin Doherty†, Bashar Nuseibeh§, Liliana Pasquale*
* Lero@University College Dublin † Lero@Trinity College Dublin ‡ Lero@University College Cork § The Open University

**Abstract**—Security attacks are rising, as evidenced by the number of reported vulnerabilities. Among them, unknown attacks, including new variants of existing attacks, technical blind spots or previously undiscovered attacks, challenge enduring security. This is due to the limited number of techniques that diagnose these attacks and enable the selection of adequate security controls. In this paper, we propose an automated technique that detects and diagnoses unknown attacks by identifying the class of attack and the violated security requirements, enabling the selection of adequate security controls. Our technique combines anomaly detection to detect unknown attacks with abductive reasoning to diagnose them. We first model the behaviour of the smart home and its requirements as a logic program in Answer Set Programming (ASP). We then apply Z-Score thresholding to the anomaly scores of an Isolation Forest trained using unlabeled data to simulate unknown attack scenarios. Finally, we encode the network anomaly in the logic program and perform abduction by refutation to identify the class of attack and the security requirements that this anomaly may violate. We demonstrate our technique using a smart home scenario, where we detect and diagnose anomalies in network traffic. We evaluate the precision, recall and F1-score of the anomaly detector and the diagnosis technique against 18 attacks from the ground truth labels provided by two datasets, CICIoT2023 and IoT-23. Our experiments show that the anomaly detector effectively identifies anomalies when the network traces are strong indicators of an attack. When provided with sufficient contextual data, the diagnosis logic effectively identifies true anomalies, and reduces the number of false positives reported by anomaly detectors. Finally, we discuss how our technique can support the selection of adequate security controls.

## 1 INTRODUCTION

The number of cyber attacks and vulnerabilities reported is on the rise [1]–[3]. Among the different types of attacks, we refer to unknown attacks collectively as novel variants of known attacks, technical blind spots [4], and previously undiscovered attacks (commonly referred to as zero-day attacks). These attacks pose challenges to providing security with a long-term protective outlook (i.e., sustainable security [5]) due to the limited number of automated techniques that diagnose them and identify adequate security controls. Although anomaly detection techniques are effective for detecting unknown attacks [6], they cannot typically reason about them [7]. Thus, security and software engineers who develop security solutions must address the challenge of diagnosing unknown attacks once detected and selecting appropriate security controls.

In this paper, we argue that one of the first steps in providing enduring security requires us to go beyond detecting zero-day attacks by diagnosing these attacks and identifying appropriate security controls. In our work, diagnosis entails identifying the violated security requirement and the class of attack that the anomaly represents. We provide a novel automated technique combining anomaly detection to detect unknown attacks at runtime with abductive reasoning to generate hypothetical diagnoses of these attacks. The choice of abductive reasoning for diagnosis is motivated by its ability to map effects to causes. This technique has long been used in diagnosis and explanation generation, mainly when provided with background information about the system under consideration [8]. Among various abductive reasoning techniques, logic-based abduction works well with partial system representations commonly used to represent modern software systems [9].

We first model the system in terms of the permitted actions and its security requirements as a logic program using Answer Set Programming (ASP) [10]. We then detect anomalies in the network traces captured from the system using an Isolation Forest (iForest) based anomaly detector and encode them in the logic program. Finally, we perform abduction by refutation [9] to identify the violated security requirement, i.e., identify which security requirements prevent a contradiction (anomaly) from existing. We use the Clingo ASP tool to model the system because it has good performance [11] and provides an expressive language to diagnose anomalies. For the anomaly detection, we chose the iForest algorithm since it exhibited the highest F1 score across most attacks within the datasets used for evaluation when compared with other widely recognised anomaly detectors such as One Class SVM and Local Outlier Factor (LOF) [12], [13].

We demonstrate our technique using a smart home scenario, where we aim to detect and diagnose anomalies in network traffic. We chose a smart home due to its dynamic nature, where changes in the devices connected to the network can potentially expand the attack surface. Cyber attacks in the home could pose risks to physical assets and users. For instance, a Denial of Service (DoS) attack against a smart lock would prevent users from locking/unlocking their doors, compromising their physical space [14].

We developed and evaluated our technique using two datasets, CICIoT2023 [15] and IoT-23 [16]. We trained the anomaly detector with unlabelled data to simulate unknown attack scenarios and used the labels only as ground truth for evaluation. We evaluated the anomaly detector's precision, recall and F1-score and the diagnosis technique against 18 attacks from the ground truth labels provided by two datasets. Our experiments show that the anomaly detector effectively identifies anomalies when the network traces are strong indicators of an attack. It also outperforms other techniques, such as Random Forest, commonly used to detect unknown attacks. When provided with sufficient contextual data, the diagnosis logic effectively identifies true anomalies, and reduces the number of false positives reported by anomaly detectors. It also outperforms explainable AI (XAI) techniques commonly used in intrusion, malware, phishing, and botnet attack diagnosis. Combining the modelling of smart homes and their security requirements with the detection and diagnosis of unknown attacks allows us to support security/software engineers in selecting security controls that could mitigate the occurrence of unknown attacks.

This paper is organised as follows. In Section 2, we survey existing literature and relevant works. Section 3 provides a detailed description of the proposed technique, while Section 4 evaluates its performance. Section 5 discusses our findings and the limitations of our work. Section 6 concludes the paper.

## 2 RELATED WORK

Unknown attacks can emerge from newly discovered vulnerabilities [12], referred to as zero-day attacks. They can also manifest through unforeseen vulnerabilities, often called blind spots [4]. Blind spots may be known in the security community and have fixes, but occur due to insufficient security knowledge [17], unforeseen changes in configuration [18], dynamic operating environments [19], delayed updates [20] and end-of-life products [21]. This section discusses approaches for detecting and diagnosing unknown attacks within cyber-physical systems (CPS), encompassing smart homes, IoT security, and computer network security.

Most unknown attack detection techniques use reasoning-based methods, supervised, or unsupervised learning [6]. Only a few works use few-shot [22] and transfer learning [23] to identify unknown variants of known attacks, particularly when less training data is available.

iPSTL (inference parametric signal temporal logic) formulae have been used to model a train brake system and detect unknown attacks by automatically identifying the parameters that distinguish an anomaly from a normal value [24]. Other techniques such as zone partitioning have also been used to identify atypical causal relationships [25] in sensor data for industrial control systems. However, such reasoning-based techniques focus on single-variable data, limiting their applicability in contexts with multidimensional data, such as larger CPS like smart homes.

Supervised anomaly detection methods utilise techniques like SARIMA and LSTM for network behaviour bounds [26], LSTM-RNN for anomalous sensor values [27], GANs with LSTM-RNN for attack detection scores [28], and privacy-preserving techniques that identify anomalies in network features expressed as Gaussian Mixture Models using a Kalman Filter [29]. Federated deep learning has been used in other work to classify attacks in IoT networks but does not discuss the detection of unknown attacks [30]. However, these approaches aim to detect known attacks in the datasets used for training, and are not evaluated against previously unseen attacks. Supervised learning techniques are also used to identify malware in unknown attacks by detecting malicious C&C communication [31], applying deep learning to identify Windows malware by analysis of Windows API calls [32], or even the use of few-shot learning to detect malware using opcode frequency analysis [33]. Another technique detects zero-day web attacks by training an RNN on benign data admitted by a Web Application Firewall (WAF). It then uses neural machine translation to to differentiate suspicious HTTP traces from benign, but does not diagnose unknown attacks or discuss effectiveness against encrypted traffic [34]. Moreover, these techniques are not directly applicable to CPSs due to varying processor architectures and platform-specific binaries, which may not be readily available with proprietary firmware. In addition, none of these works discuss diagnosing the detected attacks.

A more effective approach to detect unknown attacks is to use unsupervised learning techniques [7], [35], [36]. Some works use clustering to identify new anomalies in low dimensional spaces with sparse data [7] or large amounts of network data collected from an ISP [36]. Other techniques train deep learning ensembles without labels to simulate unknowns [35]. Although they effectively detect unknown attacks, these techniques do not reason about the anomalies detected.

One work on detecting zero-day attacks in malware binaries defines a zero-day attack as having a CVE but no disclosure by a vendor [37]. This work creates ground truth of virus data, identifies malicious binaries in hosts, and then analyses their presence on the Internet using Symantec antivirus software. The authors acknowledge that the technique is better suited to detecting host-based attacks with observable malware binaries. However, checking for disclosed CVEs can support the diagnosis of zero-day attacks.

To our knowledge, the literature that diagnoses unknown attacks is limited, and abductive reasoning techniques, known for their effectiveness in reasoning about behavioural anomalies [8], hold promise. Assumption-Based Truth Maintenance Systems [11] incorporate beliefs that include abduction when propositional clauses with hard true or false classifications are insufficient. HORN clauses [11] have been used to determine the most likely explanation represented in penalty logic. However, both these techniques require modelling the world using probabilistic methods that are not directly suitable to our smart home scenario and can be reserved for future work. ASP-based representations that support partial models and counterexample generation, emerge as a promising avenue for diagnosing anomalies in cyber-physical systems [38] due to their tractability and expressive language. While abductive reasoning has different application domains, e.g. generating specifications for forensic ready systems [39] and identifying violating safety properties [9], there is a research opportunity to use it to diagnose unknown attacks.

Our approach is different from root cause analysis. NIST guidelines on managing information security risk [40] and industry resources on cyber incident response [41] describe root cause analysis as the process of identifying the underlying causes or vulnerabilities responsible for a given attack. In contrast, our approach focuses on bridging the gap between network anomalies detected at run time and the security requirements defined during the system's threat modelling process when there is no access to device internals [42]. We aim to reason about the symptoms of an attack, i.e., how they manifest in the home network. We argue that identifying the underlying vulnerabilities for the network anomalies detected requires access to firmware or backend software, which we would like to explore in future work.

## 3 UNKNOWN ATTACK DETECTION & DIAGNOSIS

Our technique comprises three key activities, illustrated in Figure 1 in execution order. (1) *System Modelling*: We first create a smart home model representing possible legitimate actions and its security requirements using Answer Set Programming (ASP) (Section 3.1). (2) *Attack Detection*: We then explore and implement unsupervised machine learning to model benign device behaviour and detect network anomalies in the smart home (Section 3.2). The significance of the anomaly detection algorithm lies in its critical role in identifying abnormal network behaviour, potentially indicative of unknown attacks. (3) *Attack Diagnosis*: We represent the detected anomalies in ASP and diagnose them using abductive reasoning in refutation mode (Section 3.3). These three activities allow us to detect and diagnose unknown attacks and select adequate security controls to mitigate them.

We require data sets that represent real devices to design the smart home model and train the anomaly detector on the normal behaviour of the devices in the smart home. However, only a few smart home datasets contain real network data and/or sufficient features that can be used to reason about detected anomalies [12]. Among those available, we have selected the CICIoT2023 [15] and the IoT-23 dataset [16]. CICIoT2023 [15] includes both benign and attack data from real-world devices, providing a realistic representation of the behaviour of the smart home network. The IoT-23 dataset [16] complements the CICIoT2023 dataset by offering simulated attacks using IoT devices. Both data sets employ a flat network topology in which devices are directly connected to the router or through a network switch.

### 3.1 System Modelling

This section describes how we model the smart home and its security requirements. We require suitable language and tooling to model the system's actions and the security requirements that govern them, evaluate the satisfaction of those requirements, and reason about the anomalies detected. Answer Set Programming (ASP), a declarative programming paradigm where the system is represented by a logic program rather than its control flow, meets those needs [38]. For dynamic systems such as a smart home, which may not always be possible to model completely, ASP is a suitable candidate for modelling its actions and the rules that govern its secure operation while enabling reasoning when those rules are violated. We have chosen the Clingo ASP tool [10] for our implementation due to its expressivity and tractability [11]. We manually created the model for this work, but in future work, we intend to explore the usage of inductive learning and neuro-symbolic learning to infer the model at run time from the network traces of the system [13].

#### 3.1.1 Smart Home

We aim to represent a typical smart home network characterised by a router connected to a number of devices. An example of this topology is provided in Figure 2 in the paper that describes the CICIoT2023 dataset [15]. The smart home devices are the assets to be protected and are defined as a *type* in Clingo. The type is then used to define the domain of objects, such as the individual devices (e.g., *router*, *alexaechodot*) and relationships between them, as shown in Listing 1. Note that lines starting with '%' represent comments.

```
% Define Types
type(device).

% Devices within a smart home
device(router; alexaechodot; amazonplug;
    amcrestcamera; dlinkcamera;
    philipshuebridge; techkinlightstrip;
    irobotroomba; rpi).
```
Listing 1: System Assets

We model network interactions using a predicate, as shown in Listing 2. A predicate is a function or relation that takes one or more arguments and evaluates to true or false depending on whether the arguments satisfy certain conditions.

```
{ communicate(S,D,T,P,F) : endpoints(S),
    endpoints(D), protocols(P), S != D,
    packet_rate(F) } = 1 :- T = 0..23.
```
Listing 2: Communicate Predicate

The arguments of the predicate include the source of the network trace ($S$), the destination ($D$), the time at which it was sent ($T$), the protocol used to communicate ($P$) and a descriptor of whether the packet rate was within the learned normal limit for the device ($F$). We can model the communication in the smart home without listing every instantiation of the predicate ourselves using a choice construct ('{}'). This syntactic element enumerates the list of possibilities of a predicate. Clingo expands this construct to all possible combinations of the communicate predicate given the constraints on its variables. Clingo does not natively support temporal logic, so we discretise the time ($T$) between 0 and 23 hours. While a more granular approach can be taken to represent the time of the network flows, this representation was sufficient for our example.

We did not represent the internal behaviour of the smart home devices since we assumed that our security solution could only view the network traffic in the smart home. This is a reasonable assumption considering that it is not easy to monitor the internals of the IoT devices, which are based on proprietary firmware [42].

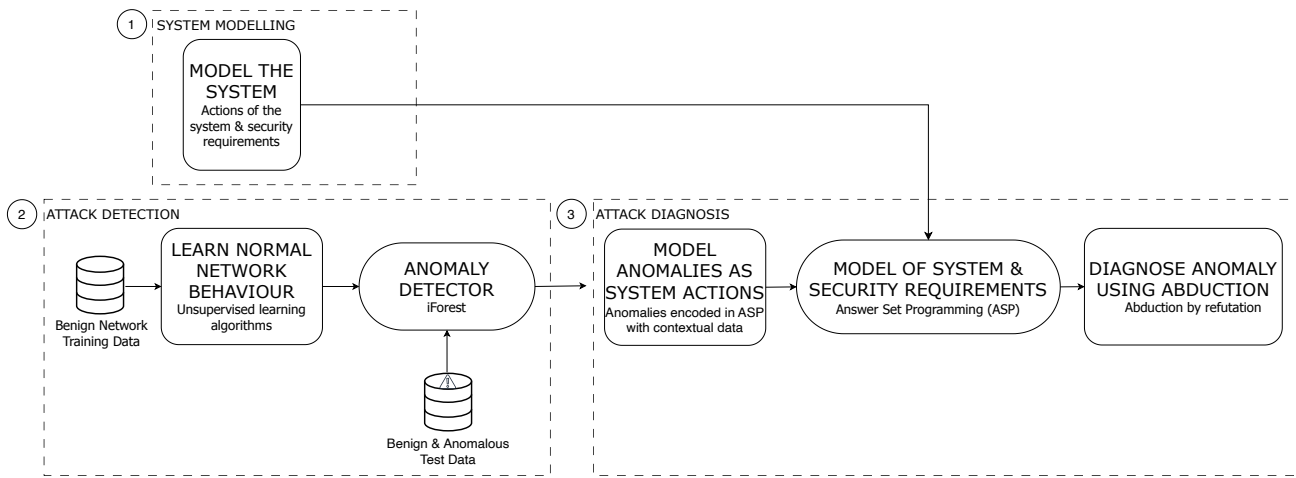

Fig. 1: Overview of the unknown attack detection and diagnosis technique

We model the domain assumptions of the system, e.g., *"the home's router is secure"*, as grounded atoms, which are predicates where the variables are replaced with constants. The assumption that a router is secure if the password is over eight characters long and uses an encrypted protocol (*wpa2*) is formalised, as shown in Listing 3.

```
% Domain assumption
password(router, 8).
encrypted(router, wpa2).
#const l = 8.

protected(router) :- password(router, L),
    L >= l, encrypted(router, wpa2).
```
Listing 3: Domain Assumptions

The model of the system is not meant to exhaustively cover all possible actions within the home that are not essential to demonstrate the effectiveness of our technique. The complete model used in our study is provided in the replication package.

### 3.1.2 Threats and Security Goals

We begin by performing a threat model of the smart home using a combination of the OWASP Threat Modelling Process [43] and an asset-centric approach [44] that assesses a home's assets, their security goals & requirements, and the threats against the system. We first identify the system's assets, which are the home's devices. We began with a simple example of a smart home but expanded it to include the list of devices in the data sets (Table 1) to enable validation. We assume that only the network traffic is observable, not the internals of the devices, which may be using proprietary firmware. We also assume that the router is secure and that the user does not intentionally communicate with malicious actors. The attack vector considered is the smart home network, and we also include one example to demonstrate physical attacks (ultrasonic voice command attack).

We then defined the system's security goals in terms of confidentiality, integrity, and availability of the identified system assets [45]. To do so, we refer to a recent taxonomy of cyber-physical threats against a smart home [18] that classifies 24 types of attacks based on violated security goals, attack vectors, and impact on the system and domestic life. These goals are listed in Table 2. The confidentiality goals of the system aim to safeguarde the secrecy of data transmitted to/from a device, and protect sensitive device data that attackers can probe to identify vulnerabilities (rows *CCOM1* & *CDEV1*). The integrity goals require trustworthiness of communication and commands, authorised inter-device communication, and the integrity of device software and firmware (*ICOM1, ICOM2, & IDEV1*). Lastly, the availability goals ensure that the devices remain accessible and operate as intended (*ADEV1* & *ADEV2*).

### 3.1.3 Security Requirements

Security requirements establish constraints on the system that operationalise one or more security goals [46]. Unlike security goals, requirements are verifiable and outline specific actions to be prevented, constraints on functional requirements, and assumptions regarding the system. The security requirements that operationalise the security goals discussed in Section 3.1.2 are listed in Table 2. The list of security requirements is not meant to be exhaustive but is detailed enough to demonstrate the technique's effectiveness.

The *CCOM1* requirement prevents devices from communicating using unencrypted protocols, violation of which could indicate a security misconfiguration or compromised device. *CDEV1* prevents the traffic rate of the same type from exceeding a learned limit. The attacks in the datasets that violate this requirement are Port Scan, Brute Force, and Botnet attacks (which typically begin by probing for device vulnerabilities). Brute force attacks and Port Scan attacks (i.e., reconnaissance) [15] probe a system for weaknesses and are characterised by abnormal traffic volumes to a device, similar to a DoS attack. Brute Force attacks [15] often consist of repeated submissions of requests to gain unauthorised access to confidential data of a system, such as a case with dictionary attacks that attempt to guess a password.

The *ICOM1* requirement prevents devices from communicating with malicious endpoints. The DNS Spoofing, Uploading Attack, and the C&C Communication of Botnet attacks violate this security requirement. An uploading attack [15] is one where an attacker attempts to upload a malicious file onto a device, which may violate multiple security goals and requirements. If the malicious file is a known malware binary, its signature may be available in malware databases. Without observing the malware binaries or changes to the device's internal state, these attacks are difficult to diagnose from network traces alone. Further, deep packet inspection can be difficult if the attacker uses HTTPS or some encrypted communication protocol. Given our limitation of only being able to observe the network traffic, we can still diagnose uploading attacks by their communication with a malicious endpoint. In a DNS Spoofing attack, the attacker corrupts the local DNS cache to redirect requests to malicious sites [15]. However, while it might not be possible to detect the corruption of a DNS cache from the network traffic alone, this man-in-the-middle (MitM) can be diagnosed when the traffic from a device is redirected to a malicious endpoint, similar to uploading attacks.

The *ICOM2* requirement prevents devices from communicating with each other unless authorised to do so. A stealthy Recon attack might violate this requirement. While there are valid situations in which devices may communicate with each other (e.g., a smart speaker controlling devices in a home), such communication may be an indicator of a malicious/compromised device. The datasets that we considered did not include this type of Recon attack.

The *IDEV1* requirement prevents devices from communicating outside the permitted operating hours. It is violated by cyber-physical attacks such as ultrasonic voice command attacks where the attack is carried out through a physical medium and maybe undetectable by a network monitor. The impact of this attack can be observed through unauthorised actuation of another device within the home. This attack is not present in the datasets used for our study, but we have included it to illustrate how a user-specifiable requirement can be used to diagnose an attack performed using a physical medium that is not monitored.

Requirements *ADEV1* and *ADEV2* prevent the rate of traffic to/from a device from exceeding a learned threshold. The former is violated by DoS attack whereas the latter is violated by DDoS/Botnet attacks. A denial-of-service (DoS) attack is a flood of network traffic that overwhelms the target so that it cannot receive or transmit data, i.e., renders it unavailable [18]. A distributed denial-of-service (DDoS) attack is similar to a DoS attack except that the perpetrators are typically more than one [18].

Botnet attacks can be executed in multiple stages and may violate different security requirements at each stage. For instance, communication with a C&C server can violate requirement *ICOM1*, the DDoS violates requirement *ADEV2*, and reconnaissance activities can violate requirement *CDEV1*.

The security requirements are encoded using integrity constraints, i.e., rules or conditions that must be satisfied for a model to be considered valid. If an anomaly violates an integrity constraint, it is found to violate that security requirement. For instance, Man-in-the-Middle and some

TABLE 1: List of devices and attacks

| Device | Attack |
|---|---|
| Philips Hue Bridge | DDoS HTTP Flood |
| iRobotRoomba | DNS Spoofing |
| AmcrestCamera | DoS HTTP Flood |
| DlinkCamera | DoS HTTP Flood |
| AlexaEchoDot | Mirai UDP Plain |
| AmazonPlug | Recon Port Scan |
| TechkinLightStrip | Recon Port Scan |
| Raspberry Pi | Upload Attack |
| Smart Speaker | Ultrasonic Voice Command Attack |
| Raspberry Pi | Mirai Botnet |
| Raspberry Pi | Torii Botnet |
| Raspberry Pi | Trojan Botnet |
| Raspberry Pi | Gagfyt Botnet |
| Raspberry Pi | Kenjiro Botnet |
| Raspberry Pi | Okiru Botnet |
| Raspberry Pi | Hakai Botnet |
| Raspberry Pi | IRCBot Botnet |
| Raspberry Pi | Muhstik Botnet |
| Raspberry Pi | Hide&Seek Botnet |

Malware may force a device to communicate with malicious endpoints. A security requirement that prevents such unsafe communication is encoded, as shown below.

```
:- communicate(_,X,_,_,_), malicious_endpoints
   (X).
```
Listing 4: Security Requirement as an Integrity Constraint

The '_' represents variables that are unused or irrelevant to the integrity constraint, and can take any value the predicate permits. In this case, the integrity constraint prevents any communication where the destination is a malicious endpoint. [1]

## 3.2 Detecting Anomalies in Smart Homes

To select the anomaly detection algorithm for our approach, we considered three algorithms that have been shown to be effective [12], [13] - One Class SVM (One-SVM), Local Outlier Factor and Isolation Forest. We used the Scikit-learn library [47] for our implementation due to its ease of use and our familiarity with it. We initialised the One-SVM with an RBF kernel since the network traffic in the dataset is not linearly separable [48]. For the Local Outlier Factor, we maintained the k_neighbours parameter to its default value (20) because any value above 10 could remove statistical interference [49]. We retained the default parameters of the Isolation Forest since, during our experiments, they allowed the algorithm to achieve the highest F1 score overall across all the attacks considered in the study (Tables 3 & 5).

The default predict functions in Scikit-learn use static thresholds set in the libraries, which performed poorly at identifying anomalies in our experiments. To mitigate this issue, we adopted a two-stage thresholding approach [50], in which the anomaly scores output by the algorithms are subjected to a univariate analysis technique [51], [52] that identifies thresholds for anomalous packets. We evaluated well-known univariate techniques [53] such as IQR/Tukey Fences, 95th and 99th Percentile, and Z-Score thresholds, as shown in Table 4, and selected the Z-score thresholding method since it had the best performance. A detailed discussion of the results is described in Section 4.

---

1. The complete formalisation of the security requirements is provided in the model available in the supplementary material.

TABLE 2: List of security requirements

| Index | Security Goal | Security Requirement | Attack | Diagnosis |
|-------|---------------|---------------------|--------|-----------|
| CCOM1 | Confidentiality - Communication | Devices do not communicate using unencrypted protocols | Security Misconfiguration / Vulnerability | Vulnerability/Malware |
| CDEV1 | Confidentiality - Device Data | Rate of requests of the same type do not exceed learned limit | Port Scan, Brute Force, Botnets (Mirai, Torii, Trojan, Gagfyt, Kenjiro, Okiru, Hakai, IRCBot, Muhstik, Hide&Seek) | Recon/BruteForce |
| ICOM1 | Integrity - Communication | Devices do not communicate with malicious endpoints | DNS Spoofing, Uploading Attack, Botnets (Mirai, Torii, Trojan, Gagfyt, Kenjiro, Okiru, Hakai, IRCBot, Muhstik, Hide&Seek) | MiTM/Malware |
| ICOM2 | Integrity - Communication | Devices do not communicate with each other unless authorised | Port Scan / Reconnaissance Attack | Recon |
| IDEV1 | Integrity - Device Firmware | Devices do not communicate outside permitted hours | Ultrasonic Voice Command Attack | Vulnerability/Malware |
| ADEV1 | Availability - Device | Rate of traffic to/from a single source does not exceed the learned threshold | DoS HTTP Flood | DoS |
| ADEV2 | Availability - Device | Rate of traffic to/from multiple sources does not exceed the learned threshold | DDoS HTTP Flood, Botnets (Mirai, Torii, Trojan, Gagfyt, Kenjiro, Okiru, Hakai, IRCBot, Muhstik, Hide&Seek) | DDoS |

We assume that the smart home is secure during the training phase. Thus, we only use benign data to learn the system's normal behaviour, but we use the labelled data during the test phases. This simulates a situation where the attack traces are injected into the system, since they are only seen by the model for the first time during testing. However, we note that since the datasets chosen in our study do not indicate when an attack has failed, every attack trace is assumed to be successful. This can be remedied in future work by using a test bed that includes successful and failed attacks. We created anomaly detection models per device trained on benign data and tested them using anomalous data. This was because previous studies have shown the difficulties in identifying thresholds and creating a single model for a heterogeneous collection of devices [6], [42], which could also be intermittently connected to the network. Our findings also corroborate the need to create anomaly detection models for each device.

Since the abductive reasoning logic uses ASP, the network anomalies reported by the machine learning algorithm must be represented in this language. To do that, we programmatically convert the network trace flagged as anomalous to the communicate predicate (Listing 2). The anomaly detector provides a binary classification output by considering all the features in the data set as an input. Since the communicate predicate does not require all of these features, we process the network trace in a pandas dataframe to identify the source, destination, and whether the traffic exceeds normal thresholds for this device and represent it in the format shown in Listings 5, 6, 7, and 8. The anomalous network traces are represented as atoms in the model, and we classify them as security anomalies if they are found to violate the model, e.g., a smart bulb that communicates with a Command & Control (C&C) server violates the security requirement *ICOM1*.

```
% Anomaly Trace: Device communicates with
% malware site
communicate(bulb,c2c_server1,_,https,_).
```

Listing 5: Anomaly Trace

We also introduce other contextual data (e.g., the availability of devices and the reputation of the initiator of communication) that may help diagnose the anomaly. We use the impacts on the system (e.g. malicious network traffic) and domestic life (e.g. device availability) [18] to represent contextual factors that a network monitor or smart home user can provide about an attack that can help with diagnosis. We augment the anomalies detected with three types of contextual data: (1) the computed variation from normal network traffic rate, (2) the number and reputation of the source(s) of the traffic, and (3) the availability of the device after the attack. Many types of attacks exhibit anomalous packet rates, and we use Tukey Fences, to identify the bounds of normal traffic and then identify abnormal rates [54]. If an anomalous network trace exhibits an abnormal packet rate, the *packet_rate* field of the communicate predicate is changed to "exceeds_limit". The trace below was generated for a Port Scan attack against the Amazon Plug.

```
available(amazonplug).communicate(rpi,
    amazonplug,10,https,exceeds_limit). %Recon
```

Listing 6: Contextual Factor - Rate Limit

Known malicious endpoints can be identified using IP blacklists and IP reputation checkers [55], which can be useful in identifying botnet attacks and multiple types of malware. While such checkers might miss new malware endpoints or local ones (as with the datasets we used in which the attackers were present in the LAN), a more sensitive diagnosis logic might choose to flag a locally originated packet or one with an unknown reputation as malicious. This could increase the false positive rate but lead to fewer missed anomalies. We change the *source* variable of the communicate predicate to "malicious_endpoint" when communication is done with insecure endpoints. The following trace was generated when the Raspberry Pi communicated with a C&C server during the Kenjiro botnet attack.

```
available(rpi-17-1).communicate(
    malicious_endpoint,rpi-17-1,10,https,
    within_limit). %MitM/Malware
```

Listing 7: Contextual Factor - Endpoint Reputation

To distinguish between DoS and DDoS, we modify the *source* of the communicate predicate to "single_endpoint" or "multiple_endpoints". We verified this by inspecting the network packet captures provided with the CICIoT2023 dataset. Knowing whether a device is offline after detecting an anomaly can also be useful in diagnosis. This is observable through network monitors or user input and may not be available in existing datasets. The following trace was generated after detecting the DoS HTTP Flood attack against the Dlink Camera. Note the addition of the device's availability at the beginning of the trace and the modified source.

```
not available(dlinkcamera).communicate(
    single_endpoint,dlinkcamera,10,https,
    exceeds_limit). %DoS
```

Listing 8: Contextual Factor - Availability & Number of Sources

In practice, contextual facts are relatively easy to monitor at run time since multiple tools and techniques exist to identify them. However, since we are working with a dataset and not a testbed, we have programmatically added the contextual factors ourselves while encoding each anomalous network trace in ASP, with knowledge obtained from inspection of the network traces of each anomaly and the labels of each attack.

### 3.3 Diagnosis using Abductive Reasoning

Since we are first identifying anomalous network behaviours and then diagnosing them, our objective is to identify a plausible explanation (cause) for the observed network anomalies (effect). Specifically, given a partial model of the system and anomalous observation(s) that may signify unknown attacks, we need a technique that can identify which security requirement is violated by the attack.

Abductive reasoning is a technique that identifies a plausible explanation for an event based on a given system description [8], and proves to be an apt choice for this task. We opted for it due to its efficacy in generating diagnoses in the desired format and its demonstrated tractability using Clingo ASP.

Our approach draws inspiration from prior work employing abduction by refutation to identify safety property violations in a system with a partial model [9]. The similarities with our use case are that we aim to generate explanations for anomalies (safety violations in their case), we also have a partially modelled system, and we require a solution that always terminates to be able to respond to security attacks. Abduction by refutation is a technique that identifies the conditions under which the existence of a counterexample is possible (a contradiction) given the system invariants. In our technique, given a network anomaly (counterexample) that violates the system invariants (security requirements), it means identifying which security requirement must be excluded for the anomaly to satisfy the system model (contradiction). By making the anomaly satisfy the model, we recognise the condition in which that contradiction can exit, thus identifying the violated security requirement. Our technique is described in Algorithm 1. Our model defines the security requirements for the different asset types (Listing 1) rather than for each device individually. When provided with a network anomaly, Clingo grounds every predicate for each type of device to identify which security requirement has been violated. For example, the step that identifies the counterexample of the Mirai UDP Plain botnet's DDoS phase evaluates 226,290 rules, 33 choice constructs, 47,234 atoms, and 44,944 bodies. In the absence of ASP, this would be a cumbersome and error-prone process that involves defining the predicates of normal operation and the security requirements for each device.

---

**Algorithm 1** Abduction by refutation for diagnosing network anomalies

**Input:** Model of a smart home, security requirements and network anomalies encoded in Clingo ASP

**Output:** Class of an attack and violated security requirement

**for** each anomaly **do**
    check satisfiability of the model
    **if** not satisfied **then**
        **for** requirement in list of security requirements **do**
            exclude requirement
            check satisfiability of the model
            **if** satisfied **then**
                requirement is the diagnosis
            **end if**
        **end for**
    **else**
        not a security anomaly
    **end if**
**end for**

---

While the effectiveness of our technique against the attacks in the datasets considered is discussed in Section 4, it is important to note that it applies to various cyber-physical attacks where the attack vector or impact extends beyond the cyber space [18]. Although we do not formally apply frameworks that identify advanced persistent threats such as the Cyber Kill Chain [56], the IoT-23 dataset contains attack traces from the different phases of each botnet attack, from Reconnaissance, to Command&Control, and finally Action on Objectives (i.e., DDoS exploit). We are able to diagnose the different stages of the botnet attacks, but since the dataset discretises each of these phases into different network flows and lacks temporal data for the different phases, we cannot string the diagnoses together. In this section, we demonstrate the diagnosis technique using two examples from the list of attacks evaluated. In Section 4, we discuss the diagnosis of all the attacks chosen in this study, which are among the most common attacks against smart homes [18].

The first example describes the diagnosis of the Ultrasonic Voice Command attack against a smart speaker. Devices such as smart speakers may be controlled by physical media such as voice that introduce new attack vectors. For example, some smart speakers are vulnerable to an ultrasonic voice command attack [57], [58] that allows unauthorised actuation and, in turn, control of other devices within the home through inaudible commands. Traditional security solutions that monitor only the network may miss such attacks but can potentially detect their symptoms, such as the anomalous actions initiated by the attack. Examples of such actions would be commands sent from the smart speaker to other devices outside of regular operating hours or if the user observes anomalous smart speaker activity. Although the latter requires user intervention, the former may be observable if the smart speaker sends actuation commands within the Local Area Network (LAN) [59], [60]. As an example, we included an integrity constraint to the model that identifies any commands sent by the smart speaker outside of the normal operating hours set by the smart homeowner as a violation of the integrity of the smart speaker firmware potentially due to a compromise as detailed in Listing 9.

```
% User Generated Requirement: The Smart
    Speaker must not be operated between
    23:00-04:00 hours
permitted_operating_time(T) :- T > 4, T <=
    22, T = 0..23.
:- communicate(X,_,T,_,_), X = smart_speaker
    , not permitted_operating_time(T).
```
Listing 9: Security Requirement - Communication Outside Permitted Hours (IDEV1)

Using a smart home outside the selected operating hours violates the above security requirement and results in an unsatisfiable model. This anomalous usage is represented below.

```
communicate(smart_speaker,trusted_app_server
    ,23,https,within_limit).
```
Listing 10: Anomaly Trace - Unauthorised Actuation

The abductive reasoning logic inserts this anomaly as an atom (where values replace the variables of the predicate) into the system, checks model satisfiability, and, if not satisfied, identifies the violated security requirement using Algorithm 1. The diagnosis is then output in the format shown below:

```
Violated Security Requirement:  User
    Generated Requirement: The Smart Speaker
     must not be operated between
    23:00-04:00 hours
Diagnosis: Vulnerability/Malware
```
Listing 11: Diagnosis - Ultrasonic Voice Command Attack

This diagnosis can inform the smart home owner that the smart speaker is being operated by an unauthorised user (row IDEV1 in Table 8). While there are only a few mitigations to such attacks, this diagnosis enables smart home owners and software engineers to implement an adequate security control. One of them would be to enable voice matching on the device [57]. In this case, such mitigation would require adequate explanations to the homeowner and mechanisms for user intervention, both of which we reserve for future work. However, in this example, we wish to highlight the benefit of providing a diagnosis for the anomalies observed.

The second example demonstrates the diagnosis of the DDoS HTTP Flood attack against the Philips Hue Bridge. The security requirement that prevents a DDoS attack is indicated in row *ADEV2* in Table 2, and is encoded as the integrity constraint shown below.

```
% Availability Security Requirement : Volume
     of traffic from multiple sources does
    not exceed learned threshold
:- communicate(X,Y,_,P,F), X =
    multiple_endpoints, device(Y), protocols
    (P), F = exceeds_limit, not available(X)
    . % Diagnosis: DDoS/Botnet
```
Listing 12: Security Requirement - Excess Rate of Traffic from Multiple Sources (ADEV2)

The anomaly trace generated by our technique is represented below.

```
not available(philipshuebridge).communicate(
    multiple_endpoints,philipshuebridge,10,
    https,exceeds_limit) . %DDoS/Botnet
```
Listing 13: Anomaly Trace - DDoS Attack

As described in the previous example, the abductive reasoning logic inserts this anomaly into the system, and Algorithm 1 generates the following diagnosis.

```
Violated Security Requirement:  Availability
Security Requirement : Volume of traffic
    from multiple sources does not exceed
    learned threshold
Diagnosis: DDoS/Botnet
```
Listing 14: Diagnosis - DDoS

DDoS/Botnet attacks are typically performed in multiple stages and this diagnosis helps us identify that the anomaly is caused by the botnet actuation phase which sends a large volume of traffic targeted at a device. This informs our choice of a suitable security control, which is to rate limit the traffic that is sent to a device (row ADEV2 in Table 8).

A list of the possible security controls for each of the attacks included in our study is provided in Table 8. These were identified from the works previously cited defining each attack, and from a review of the grey literature.

## 4 Evaluation

Our evaluation aims to assess the effectiveness of the detection and diagnosis techniques separately since they are different actions that must be performed when an attack occurs. For the attack detection technique, we evaluate the anomaly detector's ability to detect network attacks while minimising false negatives (i.e., Recall). For the attack diagnosis technique, we evaluate the correctness of the abductive reasoning logic in identifying the class of attack and violated security requirement (i.e., Precision). The rationale is that the anomaly detector can report a reasonable number of false positives to catch as many attacks as possible since the diagnosis technique will subsequently attempt to reason about the anomalies.[2]

2. The source code of the experiments, models and data sets used are available in the supplementary material.

### 4.1 Data Sets

In a recent survey of 44 IoT and IIoT security data sets [61], the researchers found that the most common attack types are reconnaissance (e.g., port scans), (D)DoS, MiTM (e.g., brute force), and botnet attacks. They also highlighted that most data sets feature TCP- or UDP-based traffic and noted a lack of data sets containing the ZigBee, CoAP, and other IoT/IIoT-specific protocols. For our study, we considered several options, including UNSW-NB15 [62], Edge-IIoTset Cyber Security Dataset of IoT & IIoT [63], IoT-23 [16], CICIoT2023 [15], and Ghost IoT [64], to identify those that comprehensively represent the common attack types against IoT/IIoT networks [61], and smart homes [18].

The UNSW-NB15 dataset was generated in the Cyber Range Lab at UNSW Canberra using the IXIA PerfectStorm tool across a network of routers, servers and computers. Although valuable for network intrusion detection, it lacks data from smart home devices. The Edge-IIoTset dataset includes sensor data and simulated attacks on a small set of IoT devices, which makes it more applicable to analyse sensor anomalies during attacks than general network intrusions. Ghost IoT provides a unique methodology for extracting data sets from IoT devices. Still, the data set itself is very small and does not include labels of attack traces required for evaluation.

The IoT-23 dataset, however, captures network traffic from different phases of 10 botnet attacks on smart home IoT devices, making it an excellent choice for studying botnet behaviour. In addition, the CICIoT2023 dataset includes traffic from 105 different IoT devices, simulating 33 types of attacks, and is well suited for investigating the varied device types found in typical smart homes. Thus, we chose IoT-23 and CICIoT2023 for our validation, as they represent a comprehensive range of attack types and IoT devices, providing realistic and varied data for our analysis.

To simulate unknown attacks, the attack labels in both data sets were excluded while training the anomaly detection model and implementing the abductive reasoning logic but were included during the evaluation. The CICIoT2023 dataset contains data from 33 attacks from 7 categories (e.g., DoS, Port Scan) performed against 105 real devices, but we chose one device from each category (e.g., Lighting, Home Automation) for our experiments since not all attacks were performed against all devices, and that it was sufficient to demonstrate the effectiveness of our technique. The IoT-23 dataset contains data from 14 different types of botnet attacks, from which we excluded the Linux Mirai and Hajime botnets because they are platform-specific, whereas evaluating other attacks gives better cross-platform validation. In this case, the attacks are performed by Raspberry Pi (RPi) devices within the network, and in contrast to the CICIoT2023 dataset, the devices are the perpetrators of the attacks rather than the victims.

### 4.2 Evaluation Metrics

Given the unbalanced nature of the data in an anomaly detection problem, we apply the three-sigma rule, which states that for a normal distribution of data, about 99.7% of the data falls within two standard deviations ($3\sigma$) of the mean. Consequently, we chose a conservative split of 95% benign to 5% anomalous traffic ($2\sigma$), assuming that most of the traffic in a smart home is normal, with sporadic malicious activity. In this case, the accuracy metric would not be suitable for evaluation since mispredicting every anomaly would still result in a score of 95%. Instead, we have chosen the Area Under the Curve - Precision Recall (AUC-PR), Precision, Recall and F1 score metrics to evaluate our model. The AUC-PR metric measures reparability or how well the model can distinguish between benign and anomalous traffic. Precision evaluates the proportion of anomalies reported as true anomalies, i.e., a measure of the reported false positives, while recall evaluates the fraction of true anomalies identified, i.e., a measure of the reported false negatives. The F1 score evaluates the balance of precision and recall in predicting anomalies. Since diagnosis can only occur once the attack has been detected, the two steps are sequential and the diagnosis logic will not process any false negatives from the anomaly detector. Consequently, from a security context, the false negatives of the attack detection logic must be minimised to ensure that attacks are not missed. Although we assess the AUC-PR and F1 score for the attack diagnosis, the F1 score is more relevant because it indicates a balance between missed anomalies (false negatives) and incorrect diagnoses (false positives). Although our technique effectively diagnosed attacks in both data sets, each dataset presented unique challenges, especially in understanding and cleaning the data, as discussed later in this section.

### 4.3 Data Preparation

The CICIoT2023 dataset is highly unbalanced between benign and attack traffic with very small amount of benign data in both data sets. During our initial experiments, we attempted to train a single model to identify anomalies for all the devices in Table 1. However, this model was ineffective at detecting anomalies due to the wide distribution of network traffic features (near random) and the skewed nature of the dataset. The random spread of network data was identified using the feature importance technique employing the Random Forest algorithm, which was unable to identify which features effectively distinguished benign data from malicious. The negative Inter-Quartile Rate (IQR) lower bounds that we obtained indicated the data distribution is highly skewed (to the right), i.e., larger volumes of traffic with higher packet rate or inter-arrival time (IAT) etc., and also showed the presence of extreme outliers, and potentially a small sample size in some cases. The dataset consolidates multiple network packets into single entries in the dataset called flows. We found the number of packets non-uniform across multiple flows, yielding incomparable flow metrics. We addressed this by creating a normalized packet rate since the duration of the flow and the number of packets in the flow were available. While the CICIoT2023 paper describes all attacks being performed against all devices, we found that to be incorrect after multiple experiments which yielded no inferrable statistical correlations between benign and anomalous traffic, and upon inspection of the network packet captures (PCAP). We reached out to the creators of the dataset to find the actual list of attacks against devices and created Table 1 based on it. Consequently, this led to us regenerating the data sets from the original PCAP files since the original files consisted of a mix of benign packets for devices not being attacked with attack packets specific to each device.

To regenerate the PCAP files, we first identified the MAC addresses of the devices in Table 1 from the CICIoT2023 paper and, based on our correspondence with the authors, identified the attacks performed against those devices. We downloaded all the PCAP files provided and filtered them according to the benign and attack data for each device using the tcpdump utility. We then modified the Generating_dataset.py file provided in the supplementary material of the dataset to process the newly generated files. We used it to create the CSV files to train and evaluate the models. The tcpdump utility outputs packets captures in the PCAP-NG format, but the script required the PCAP files to be present in the pcap format, so we also needed to convert the files generated by the tcpdump utility appropriately.

The IoT-23 dataset consists of multiple files in the Zeek log format, which needed to be cleaned and converted into CSV format for processing by the iForest. We found that the labels provided for the network traffic were inconsistent, and needed to be cleaned. Some fields such as the IP source address, and source headers were removed to be consistent with the features in the CICIoT2023 dataset. The packet rate also needed to be computed from the features available. Given the large size of the dataset, we choose to process it in chunks. Given that the iForest only accepts numerical features, some values in different formats needed to be converted to a numerical format, while others such as protocol, service type and connection state were categorically encoded instead to make the data suitable for consumption by the machine learning model.

### 4.4 Attack Detection

We compared the performance of the three machine learning algorithms at anomaly detection tasks using the CICIoT2023 dataset to identify the most suitable one for our use case, i.e., one with a consistently high F1 score demonstrating a good balance between precision and recall. We used k-fold cross-validation with a parameter of 10, which has generally been effective in practice [65] and present the results for One-SVM, LOF, and iForest in Tables, 3. This also allowed us to perform a more robust evaluation that provides an opportunity for different samples in the data set to appear in both training and test data sets while also averaging the results of each iteration. While the One-SVM was the most effective at identifying DNS Spoofing attacks, and the LOF was better at identifying Port Scan attacks, the iForest showed better results overall, leading us to select it for the rest of our experiments.

To select the threshold for the anomaly scores reported by the iForest, we evaluated the IQR/Tukey fences, 95th and 99th percentile, and Z-Score threshold techniques (results shown in Table 4. The IQR/Tukey Fences thresholding method showed varying results, often resulting in low precision and recall scores. Although this method was moderately effective for specific cases (e.g., Recon Port Scan on Techkin Light Strip with an F1 score of 0.508), it could not achieve consistently good performance across the whole set of attacks targeting different smart home devices.

Percentile-based thresholding techniques (e.g., 90th, 95th, and 99th) exhibited poor performance in capturing anomalous behaviour across the dataset. The 99th percentile was a very high threshold and excluded most anomalies. Consequently, its results were excluded from our study. The 90th and 95th percentile performed better and were very similar to each other, but they could not achieve meaningful precision or recall for most attack types. This was likely due to their inability to capture subtler deviations in anomaly scores indicative of anomalous behaviour. Consequently, the F1 scores were consistently low or zero across most cases, indicating the limitations of percentile-based approaches for our use case. Since the results were very similar, for brevity, we provide the results of the 95th percentile in Table 4.

The Z-Score thresholding method demonstrated strong performance across all attack types, achieving high precision, recall, and F1 scores for most attacks. The flexibility of this method in detecting both lower- and upper-bounded anomalies makes it effective in discriminating benign behaviour from attacks. To select the Z-score thresholds, we inspected the histograms of the anomaly scores of each model. For the Isolation Forest, we found that a threshold < -0.4 is slightly less than the mean of anomaly scores, and a threshold of 0 best suits the benign data. Similarly, we chose a threshold of 0 for the One-SVM model since the histograms showed positive values as inliers. With the LOF, thresholds were more widely spread depending on the attack, but we obtained very similar results with thresholds of 0, -0.4, -0.5, and -1.0.

The iForest was consistently able to correctly classify DDoS HTTP Flood, DNS Spoofing Mirai UDP Plain and Port Scan (against the Amazon Plug) as anomalies. However, we observed lower recall, i.e., a higher number of false negatives, in the case of DoS HTTP Flood attacks against Amcrest and Dlink cameras. This could be because the attack data traffic pattern appears similar to the benign ones for both cameras since they generate a continuous stream of network traffic. Similarly, the Port Scan attack traffic appears to be very similar to the benign traffic, explaining the poor precision, i.e., high false positives, for the TechkinLightStrip device. The results with the Amazon Plug were satisfactory despite having less data available for training.

The results of the anomaly detection experiments with the IoT-23 dataset are shown in Table 5. The iForest was very effective in identifying anomalies in most cases, but some discussion on exceptions is needed. The Torii and Trojan attacks did not have adequate data to train the iForest model as indicated by the poor performance. However, the abductive reasoning logic was still effective at identifying false positives. The Okiru attack data had very few benign samples to train a model. The Linux Mirai and Hajime attacks were ignored from our study because they are platform-specific.

### 4.5 Attack Diagnosis

To evaluate the abductive reasoning technique, we modified the attack labels provided in the dataset to annotate the violated security requirements and the class of attack the anomaly could belong to. When the anomalies themselves are included in the ASP model of the smart home, we convert the anomalous network traces into ASP notation and augment it with contextual facts similar to how they would be obtained from monitoring the system or from a user of the system. This is done to simulate a situation where the attack is unknown, but we know the security requirements we would like to ensure. While augmenting anomalies with contextual factors can be viewed as similar to training with labeled data that may artificially improve the performance of the algorithm, we would like to point out that we only added contextual factors that are readily observable or verifiable, that have been studied before in academic & grey literature.

TABLE 3: Comparison of iForest, One-SVM, and LOF Results with the CICIoT2023 Dataset

| Attack | Device | iForest | | | | One-SVM | | | | LOF | | | |
|---|---|---|---|---|---|---|---|---|---|---|---|---|---|
| | | Precision | Recall | AUC PR | F1 Score | Precision | Recall | AUC PR | F1 Score | Precision | Recall | AUC PR | F1 Score |
| DDoS HTTP Flood | Philips Hue Bridge | 0.9536 | 0.9774 | 0.9764 | 0.9653 | 0 | 0 | 0.6365 | 0 | 0.8857 | 0.1575 | 0.7680 | 0.2663 |
| DNS Spoofing | iRobot Roomba | 0.7188 | 0.9978 | 0.9991 | 0.8345 | 0.9935 | 0.9978 | 0.9980 | 0.9956 | 1.0000 | 0.6483 | 1.0000 | 0.7856 |
| DoS HTTP Flood | Amcrest Camera | 0.9997 | 0.6946 | 0.9954 | 0.8197 | 0.8621 | 0.2099 | 0.4764 | 0.3346 | 0.9097 | 0.0123 | 0.7086 | 0.0243 |
| DoS HTTP Flood | Dlink Camera | 0.9998 | 0.5012 | 0.9740 | 0.6666 | 0.9994 | 0.6640 | 0.9033 | 0.7979 | 0.9998 | 0.3588 | 0.9757 | 0.5281 |
| Mirai UDP Plain | Alexa Echo Dot | 0.9994 | 0.8092 | 0.9935 | 0.8902 | 0.9423 | 0.2488 | 0.4218 | 0.3936 | 0.7930 | 0.0087 | 0.4731 | 0.0172 |
| Recon Port Scan | Amazon Plug | 0.8414 | 0.9068 | 0.9193 | 0.8726 | 0.7554 | 0.9375 | 0.8607 | 0.8364 | 0.9849 | 0.9276 | 0.9449 | 0.9553 |
| Recon Port Scan | Techkin Light Strip | 0.5002 | 0.8072 | 0.9124 | 0.6174 | 0.4498 | 1.0000 | 0.9118 | 0.6202 | 0.9691 | 0.6788 | 0.8512 | 0.7949 |
| Upload Attack | RPi | 0.5320 | 1.0000 | 0.9697 | 0.6920 | 0.3558 | 1.0000 | 0.8906 | 0.5240 | 0.0571 | 0.0364 | 0.2151 | 0.0433 |

TABLE 4: Comparison of Anomaly Threshold Methods using Isolation Forest with the CICIoT2023 Datasets

| Attack | Device | Z-Score | | | | IQR | | | | 95th Percentile | | | |
|---|---|---|---|---|---|---|---|---|---|---|---|---|---|
| | | Precision | Recall | AUC PR | F1 Score | Precision | Recall | AUC PR | F1 Score | Precision | Recall | AUC PR | F1 Score |
| DDoS HTTP Flood | Philips Hue Bridge | 0.9536 | 0.9774 | 0.9764 | 0.9653 | 0 | 0 | 0.9764 | 0 | 0 | 0 | 0.9764 | 0 |
| DNS Spoofing | iRobot Roomba | 0.7188 | 0.9978 | 0.9991 | 0.8345 | 0 | 0 | 0.9991 | 0 | 0 | 0 | 0.9991 | 0 |
| DoS HTTP Flood | Amcrest Camera | 0.9997 | 0.6946 | 0.9954 | 0.8197 | 0.0612 | 0.0033 | 0.9954 | 0.0062 | 0.0662 | 0.0035 | 0.9954 | 0.0067 |
| DoS HTTP Flood | Dlink Camera | 0.9998 | 0.5012 | 0.9740 | 0.6666 | 0.3017 | 0.0020 | 0.9740 | 0.0040 | 0.7776 | 0.0175 | 0.9740 | 0.0342 |
| Mirai UDP Plain | Alexa Echo Dot | 0.9994 | 0.8092 | 0.9935 | 0.8902 | 0.0794 | 0.0056 | 0.9935 | 0.0105 | 0.0041 | 0.0002 | 0.9935 | 0.0004 |
| Recon Port Scan | Amazon Plug | 0.8414 | 0.9068 | 0.9193 | 0.8726 | 0.6234 | 0.8000 | 0.9697 | 0.6991 | 0 | 0 | 0.9193 | 0 |
| Recon Port Scan | Techkin Light Strip | 0.5002 | 0.8072 | 0.9124 | 0.6174 | 0.8444 | 0.4001 | 0.9124 | 0.5084 | 0 | 0 | 0.9124 | 0 |
| Upload Attack | RPi | 0.5320 | 1.0000 | 0.9697 | 0.6920 | 0.6234 | 0.8000 | 0.9697 | 0.6991 | 0 | 0 | 0.9697 | 0 |

TABLE 5: Isolation Forest Results using Z-Score Threshold with the IoT-23 Dataset

| Attack | Device | Precision | Recall | AUC PR | F1 score |
|---|---|---|---|---|---|
| Mirai | RPi | 0.9704 | 0.6202 | 0.9476 | 0.7567 |
| Torii | RPi | 0.0103 | 1.0000 | 0.9237 | 0.0204 |
| Trojan | RPi | 0.0029 | 0.6000 | nan | 0.0058 |
| Gagfyt | RPi | 0.6946 | 1.0000 | 0.9842 | 0.8179 |
| Kenjiro | RPi | 0.9999 | 0.7274 | 0.8289 | 0.8421 |
| Okiru | RPi | 0.9220 | 0.0006 | 0.7305 | 0.0012 |
| Hakai | RPi | 0.9975 | 1.0000 | 0.9971 | 0.9987 |
| IRCBot | RPi | 0.9999 | 0.7500 | 0.7960 | 0.8571 |
| Muhstik | RPi | 0.9932 | 0.5178 | 0.8730 | 0.6807 |
| Hide & Seek | RPi | 0.8856 | 0.9999 | 0.9093 | 0.9393 |

In all cases, the range of benign network packet rates is computed using IQR ranges, and if any anomaly trace has an abnormal rate, the flow is marked as one that exceeds the permissible limit. DDoS HTTP Flood and botnet attacks are typically performed by multiple devices as observed in the PCAP files of the CICIoT2023 dataset, although it is not included as a feature. The contextual factor added to aid in the diagnosis is that the source of the observed anomaly contains multiple endpoints. The malware Uploading Attack, communication with C&C servers in botnet attacks, and DNS Spoofing attacks are marked as occurring with malicious endpoints. This can be verified using IP reputation checkers and checking the DNS resolution for requests. In our experiments, for some cases of the DNS spoofing attack, we identified destinations with lower reputation scores, however, the same could not be done for packets from within the LAN. In the case of the various DoS, DDoS and botnet attacks, the two data sets do not indicate if the victims of the attack went offline. We have assumed that the attacks are successful and that the devices have gone offline to aid in the diagnosis, which can easily be achieved by pinging the devices. If these attacks are performed against the devices but do not go offline (i.e., facts are wrong), we still diagnose them as Recon/Brute Force attacks since it is difficult to distinguish between them.

Since we simulate scenarios where the attacks are unknown, the diagnoses we generate are on the violated security requirement and the class of the attack that those violations indicate. In this context, we needed to relabel the attacks in the dataset to provide valuable results. We combined DDoS and Botnet attacks as a potential diagnosis since botnet attacks are often a type of DDoS attack. Due to the similarities in the manifestation of the attacks, Recon and Brute Force attacks are combined. Any communication with a malicious endpoint could indicate many different types of attacks, such as with C&C flows of botnet attacks and DNS Spoofing, and they were grouped as Man-in-the-Middle (MitM) or Malware. We used the precision, recall and F1 score metrics to evaluate the abductive reasoning technique as well and for the same reasons. The results of the abductive reasoning experiments for the CICIoT2023 are shown in Table 6, and those for the IoT-23 dataset are in Table 7.

The diagnosis technique works well against most attacks, except the DoS HTTP Flood attack against the Amcrest camera and the malware upload attack against the Raspberry Pi. Upon inspecting the ASP representation of the anomalies, we found that the IQR technique was ineffective at distinguishing normal from attack packet rates which are an important criterion for diagnosing DoS attacks. We believe this is due to the similarity between the normal data streamed from the device and the DoS attack patterns in the Amcrest camera. We do not observe the same phenomenon in the Dlink camera against which the same attack was performed. In the future, we would like to explore other temporal analysis techniques that better model the flow of network packets to determine anomalous traffic rates [26]. In the case of the uploading attack, we could not iden-tify any malicious endpoint using the previously described techniques, so we did not augment the anomalies with the data required to perform a diagnosis. To improve the malicious endpoint detection, we could perform a stricter filtering of the results of an IP reputation checker wherein all inconclusive results are considered potentially unsafe.

TABLE 6: Abductive Reasoning Results with the CI-CIoT2023 Dataset

| Attack | Device | Precision | Recall | F1 score |
|---|---|---|---|---|
| DDoS HTTP Flood | Philips Hue Bridge | 0.8348 | 0.8348 | 0.8348 |
| DNS Spoofing | iRobot Roomba | 0.8710 | 0.8710 | 0.8710 |
| DoS HTTP Flood | Amcrest Camera | 0.1205 | 0.1205 | 0.1205 |
| DoS HTTP Flood | Dlink Camera | 0.9995 | 0.9995 | 0.9995 |
| Mirai UDP Plain | Alexa Echo Dot | 0.9995 | 0.9995 | 0.9995 |
| Recon Port Scan | Amazon Plug | 0.8000 | 0.8000 | 0.8000 |
| Recon Port Scan | Techkin Light Strip | 0.8148 | 0.8148 | 0.8148 |
| Upload Attack | RPi | 0.2105 | 0.2105 | 0.2105 |

TABLE 7: Abductive Reasoning Results with the IoT-23 Dataset

| Attack | Device | Precision | Recall | F1 score |
|---|---|---|---|---|
| Mirai | RPi | 0.9933 | 0.9933 | 0.9933 |
| Torii | RPi | 0.8667 | 0.8667 | 0.8667 |
| Trojan | RPi | 1.0000 | 1.0000 | 1.0000 |
| Gagfyt | RPi | 0.8571 | 0.8571 | 0.8571 |
| Kenjiro | RPi | 0.5065 | 0.5065 | 0.5065 |
| Okiru | RPi | 0.9319 | 0.9319 | 0.9319 |
| Hakai | RPi | 1.0000 | 1.0000 | 1.0000 |
| IRCBot | RPi | 1.0000 | 1.0000 | 1.0000 |
| Muhstik | RPi | 0.9218 | 0.9218 | 0.9218 |
| Hide&Seek | RPi | 0.9625 | 0.9625 | 0.9625 |

In the case of the IoT-23 dataset, each attack consisted of multiple phases, each considered as an individual attack. For example, the Botnet attack includes a Recon stage for probing the devices, communication with the C&C server, and finally a DDoS attack. While the results in Table 7 are labelled by the class of the attack, the diagnosis metrics were computed by aggregating the individual values obtained for each stage of the attacks. Our technique is effective at diagnosing all attacks except the Kenjiro attack. Since each anomaly originated from a single source, they were falsely diagnosed as DoS, not DDoS/Botnet. This underscores the need for accurate facts for diagnosis.

The diagnosis of anomalies took between 0.33s per anomaly in the best case to 0.8s in the worst case in our experiments. While this diagnosis time is sufficient to react to an attack, there is scope for improvement.

### 4.6 Comparison with benchmarks

Our approach to sequentially perform attack detection and diagnosis is novel and we did not find any technique in the literature to directly benchmark our work against. However, the individual components of our approach can be compared with the techniques recommended in the literature.

We compared our *attack detection* technique with a machine learning algorithm popularly used to detect unknown and zero-day attacks, using the CICIoT2023 dataset since it contains data from real IoT devices and various attack types. Using the same dataset, we also compared our *attack diagnosis* technique with feature relevance and explainable AI (XAI) techniques commonly used in intrusion, malware, phishing and botnet attack detection [66]. We applied the techniques against the CICIoT2023 data set, which we have cleaned and prepared, and compare the results of each with that of our approach.

For the machine learning algorithm, we selected the Random Forest (RF) since (1) it was recommended by the CICIoT2023 dataset creators, who observed strong performance in their experiments [15], and (2) among non-deep learning (DL) techniques, RF is widely used for detecting zero-day attacks, as highlighted in recent surveys [12]. We excluded deep learning techniques because our dataset does not contain sufficient samples for all the attacks.

Our initial experiments with RF resulted in overfitting (precision, recall and F1 score = 1) since we have hugely imbalanced data. To address this issue, we first ensured there was no data leakage (between training and testing) and then tried stratified sampling to ensure that the same class balance was maintained in both training and validation. That still resulted in an overfitted model. We next applied SMOTE to rebalance the minority class and a standard scaler to reduce the impact of large outliers. Then, we trained the RF classifier, which still resulted in overfitted classes for most attacks. While these results are better than our approach, an overfitted model is not generalisable to attack traces not present in the dataset.

Supervised learning techniques, such as RF, can accurately identify the class of attack since they typically excel at classification tasks when provided with sufficient labelled data. However, they cannot be applied when the attack traces largely differ from the training data. Thus, we compared our attack diagnosis technique with feature relevance techniques, including feature and permutation importance, which explain which features contributed the most to the classification. We repeated the experiment for each device-attack pair in the dataset and included all the results in the replication package, but we discussed a few examples here.

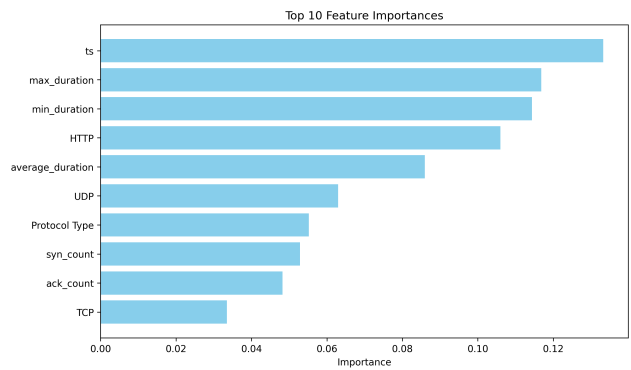

Fig. 2: Feature Importance Results of DDoS HTTP Flood attack against Philips Hue Bridge

We then applied two feature relevance techniques (Feature Importance and Permutation Importance), and two explainable AI techniques (SHAP and LIME) on the outputs of the attack detection phase, and compared the results with our attack diagnosis approach. For the feature importance experiments, we identified the top 10 important features of 62 available in the data set for each attack-device pair. Looking at the Feature Importance (FI) results of the DDoS HTTP Flood attack against a Philips Hue Bridge (Figure 2), it is unclear as to (1) what is abnormal about the timestamp (ts), (2) what the min/max/average durations of the flows should be (whether they have been exceeded or throttled), (3) why UDP is more important than TCP despite it being an HTTP based attack. Only if the type of attack is known in advance is it possible to surmise that the SYN/ACK counts could be due to multiple attempted connections, indicative of a DoS/Recon/Brute Force attack.

In addition to this, we also applied the Permutation Importance (PI) technique that attempts to identify the predictive features post-training, by assessing the performance of the model in the absence of selected features. PI was unable to identify any explanations, consistently returning a score of 0 for most features across the different devices and attacks. Since we had already attempted data rebalancing and cross-validation techniques, a search of the grey literature indicates that this is likely due to (1) small and insufficiently complex data patterns (which is the nature of these attacks), or (2) an overfitted model resulting from that dataset.

Despite having an overfitted model that should perfectly capture the dataset, neither approach provides enough information to diagnose the violated security requirement and the class of attack or allow us to select an appropriate mitigation strategy. In comparison, our diagnosis technique identifies the violated security requirements, as shown in Listing 12, and prints the diagnosis, as shown in Listing 14. This enables the selection of a mitigation strategy that prevents high traffic rates to a device and also inspects the reputation of the multiple sources sending nearly identical traffic to the device.

We then compared our attack diagnosis technique with posthoc-explainability XAI techniques. Compared with transparent-model approaches (by design), posthoc-explainability techniques (after prediction) are better for decision tree-based methods (such as RF and iForest) [67]. Of the explainable AI techniques commonly used in IDS, malware, phishing and botnet detection, SHAP and LIME are among the most commonly used for posthoc-explanations for machine learning-based approaches [66]. These two reasons prompted the selection of the SHAP and LIME techniques for our experiments. SHAP is better suited for global explanations (i.e., generalised explanations of the model prediction), and LIME is better suited for local explanations (i.e., why the model identified a particular instance as anomalous).

Compared to feature relevance techniques, which also give global explanations, SHAP values provide more comprehensive insights into how individual features contribute to the model's predictions and are better when there are complex interactions between features (as is the case with RF). The results obtained by applying SHAP for the DDoS

HTTP Flood attack against a Philips Hue Bridge are in Figure 3, and the results of applying the technique to all the attacks in our study are included in the replication package.

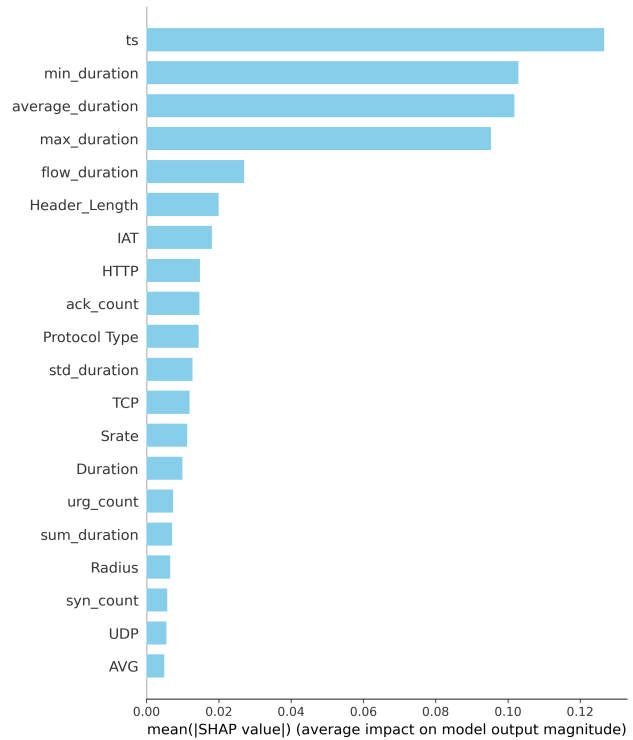

Fig. 3: SHAP Results of DDoS HTTP Flood attack against Philips Hue Bridge

Using SHAP, we obtained similar results regarding feature and permutation importance in that we could not identify (1) why features were deemed important and (2) what the normal range of values for those features looked like. Consequently, they have similar limitations to feature relevance techniques (as discussed previously) regarding their diagnosis capabilities.

In comparison to the feature relevance-based approaches, LIME is better at reasoning about individual instances while providing the range of values that impacted the prediction and which class the features contributed to. Since it provides explanations for each anomaly, we randomly sampled 10 predictions from the RF model for each attack in the CICIoT2023 dataset to study the explanations provided. The explanations for the HTTP DoS Flood attack against the Philips Hue Bridge indicate that flows containing HTTP are benign rather than attack samples despite having stratified sampling. In some cases (Figure 4), inter-arrival time (IAT) was used to predict benign flows as well when it should have contributed to the positive class (attack). In another case (Figure 5), IAT was used to correctly explain an HTTP flood and the range of values for the flow to be an attack was also provided. For mispredicted outcomes (Figure 6), the model identified HTTP traffic with higher ACK flags and IAT as attack flows, when in fact, they were benign flows similar to an HTTP DoS attack.

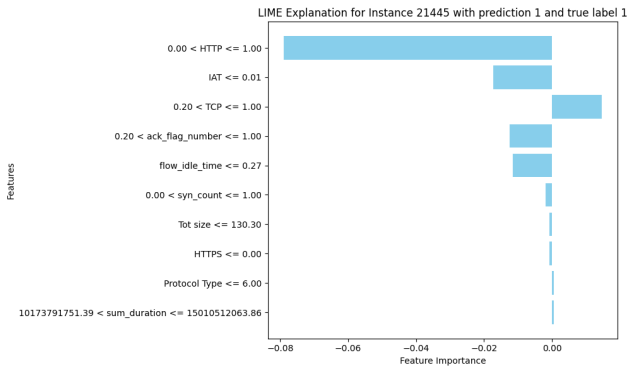

Fig. 4: LIME Results of DDoS HTTP Flood attack against Philips Hue Bridge where IAT incorrectly indicates anomalous data to be benign

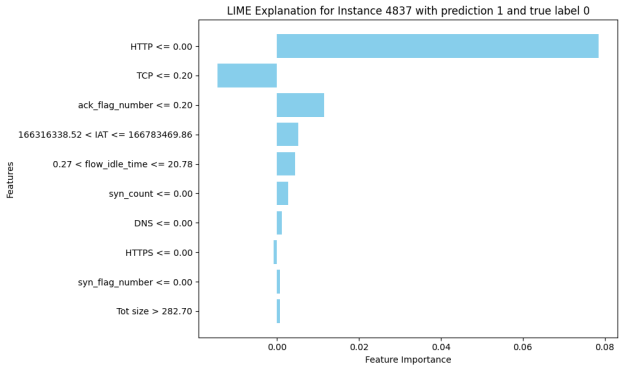

Fig. 5: LIME Results of DDoS HTTP Flood attack against Philips Hue Bridge using IAT correctly to classify attack data

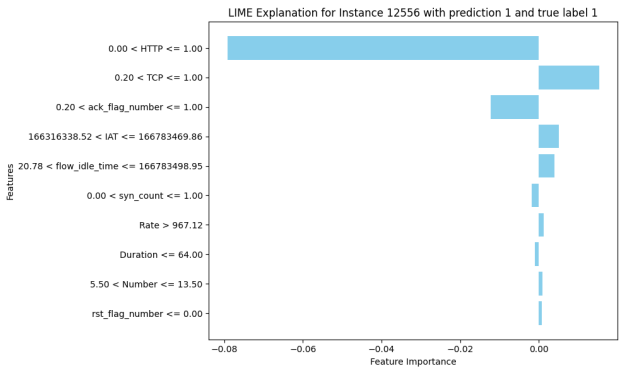

Fig. 6: LIME Results of DDoS HTTP Flood attack against Philips Hue Bridge for mispredicted outcomes

In the case of the Alexa Echo Dot against which the Mirai UDP Plain attacks were performed (Figure 7), the top 10 features ranked by importance mostly indicate an inclination to the negative class (benign), even though the prediction was correctly made for the positive class (attack). This could be due to the complex relationships between the features in the Random Forest or the aggregate of the positive features outweighing the negative ones despite individual values of the positive features being low. In this case, despite correctly detecting the anomalies, the features with the highest importance do not give useful information to diagnose the attacks. Further, since LIME is better suited to local explanations (for that particular anomalous instance), its explanations could contain features that do not contribute to the prediction.

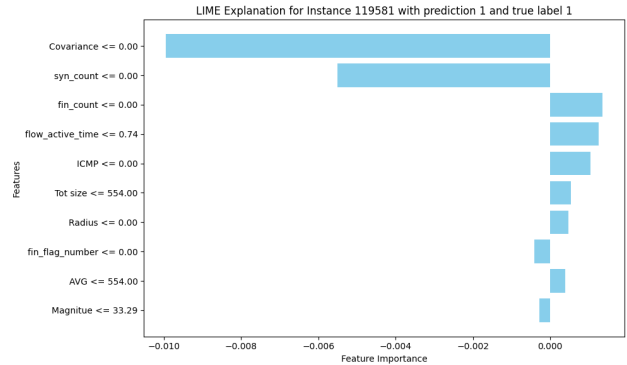

Fig. 7: LIME Results of Mirai UDP Plain attack against Alexa Echo Dot

None of these explainability techniques make use of contextual data such as rate limits, endpoint reputation, availability of devices, and number of sources (Listing 6, 7, 8) from the system in order to reason about the anomalies detected. While it might appear that contextual data could be used to train a supervised learning model, in the case of unknown attacks, which data to use might not always be apparent and would require re-training the model if discovered after deployment. For example, a flood of packets may occur due to Recon attacks or DoS/DDoS attacks. A supervised learning algorithm would not use contextual data about the availability of the targeted device. Thus, it can misclassify a DoS attack as a Recon attack even if the attack makes the targeted device unavailable.

## 5 DISCUSSION

Our approach combines behaviour- and behaviour-specification-based anomaly detection [6]. By detecting network anomalies (behaviour-based) and reasoning about them using the formalism of ASP (behaviour-specification-based), we can mitigate false positives of the anomaly detector while overcoming the need for complete system models. The expressivity of the modelling language allows easy extension of security requirements to include other user preferences or exceptions such as If-This-Then-That (IFTTT) [68] rules that the user might have configured, which are not modelled in our current work. The actual mechanisms in which user input may be sought are left for future work. In the rest of this section, we discuss how our approach can support the selection of security controls and the implications for human intervention.

### 5.1 Selecting Security Controls

Although we know and explicitly represent the security requirements that can be violated, we may not know how the attacks that violate them can materialise in specific smart home deployments. We contend that understanding the attacks in terms of the security goals and requirements they violate enables the selection of suitable security controls and eliminates the need for precise attack identification. While the automated selection of security controls after diagnosing an attack is left for future work, in this section, we provide examples of how the identification of the violated security requirements can inform the choice of appropriate mitigation strategies. In the case of a Port Scan attack, identifying the abnormal traffic enables us to better filter the network traffic sent to a device. If the perpetrator of the attack is within the home network, a network sandbox may be employed to isolate all traffic from that device. Knowing exactly the malicious behaviour exhibited by the device could enable a search within vulnerability databases to identify if the attack was known and/or mitigated. Based on this analysis, (1) the home owner could be requested to remove the device from the network, or (2) apply security updates if the vulnerability has been fixed by the vendor, or (3) even identify the trade-offs that allow continued operation of the device if the user chooses not to remove it. Some potential mitigations for the attacks included in our study are described in Table 8.

TABLE 8: List of Security Controls

| Index | Security Control |
|---|---|
| ICOM1 | Filter traffic to/from malicious endpoints |
| ICOM2 | Block network traffic between devices initiated by a device that is not whitelisted |
| ICOM2 | Network sandbox the source of suspicious traffic |
| IDEV1 | Enable voice matching features |
| IDEV1 | Block all actuation outside permitted hours |
| ADEV1 | Rate limit single-source traffic |
| ADEV2 | Rate limit multi-source traffic |
| CDEV1 | Filter traffic to/from malicious endpoints |
| CCOM1 | Configure the device to use secure communication protocols |
| ALL | Update device firmware for patches found in vulnerability databases |

## 5.2 Adapting to changes at run time

In a flat network configuration, modifications are limited to the addition and removal of devices, or the modification of device behaviour. In the case of adding a new device, it would be necessary to create a dedicated anomaly detection model for that device. However, provided the device operates over TCP/IP, the ASP model would remain unchanged. If a device's behavior is modified—due, for example, to an update or a shift in user interaction patterns—this can be addressed by (1) the user notifying the system of the update, or (2) the anomaly detector identifying a series of anomalies that do not violate the security requirements defined. This scenario necessitates retraining the anomaly detection model and may also introduce new security requirements depending on the extent of the changes.

We assume that changes in network topology are intentionally made by the smart home user and using secure devices (i.e., router, WiFi extender). When a conventional WiFi extender is used, all inbound and outbound traffic remains monitorable at the main router, allowing the approach to function without modification. In cases where a separate subnet is created within the home—either through an additional router or another WiFi extender—the approach would need to be applied individually within each subnet.

## 5.3 Implications for Human Intervention

Although our work focuses on automating the detection and diagnosis of unknown attacks using anomaly detection and abductive reasoning, it also sheds light on potential avenues for human intervention. We envision the involvement of different stakeholders: users (the homeowners or house tenants), security/software engineers responsible for securing the smart home devices or the home network, and pen testers tasked to discover new vulnerabilities by performing offensive testing. These stakeholders can support the execution of specific activities that, in our approach, cannot be automated and should be delegated to humans. They can also improve unknown attack detection and diagnosis and ultimately identify and even execute more robust security controls.

Our approach performs unknown attack detection based on the smart home network behaviour and on contextual information that can be monitored automatically: variation from normal network traffic rate, the number and reputation of the source(s) of the traffic, and the availability of the device affected by an anomaly. The presence of a user in the house opens up the possibility of monitoring some of the abovementioned contextual information (e.g., device availability) and information about the smart home device behaviour (e.g., failed Transport Layer Security (TLS) verification, unexpected pop-ups), which cannot be done automatically. This additional information has the potential to rule out false positives flagged by an anomaly detector (e.g., a DoS attack is flagged, but the targeted device is up and running). We also envision the possibility for the user to pro-actively monitor specific information that could help discover anomalies that would not be noticed from the network behaviour analysis, for example, situations when devices exhibit behaviour not instructed by the user (e.g., a smart speaker unexpectedly reproducing audible sounds, a smart light unexpectedly switching on and off) [69]. This monitored information could support the identification of anomalies overlooked by an anomaly detector.

There could be a situation when the abductive reasoning logic cannot identify a specific security requirement responsible for the anomaly. This may be because the anomaly requires a more complex diagnosis to be explained (more than one security requirement is violated simultaneously) or some domain assumptions present in the model are no longer valid. In such cases, a security/software engineer could be involved in supporting the diagnosis and identifying the classes of attacks that could have caused the anomaly and violated security requirements and domain assumptions. If a diagnosis is identified, an attacker may have exploited an unpatched vulnerability in the targeted device to cause the anomaly. A security/software engineer could help identify important security updates that should be performed on a smart home device. However, a zero-day vulnerability may be present if the device is up to date. Information about the anomaly, contextual information, and the target device could be shared with pen testers at the vendor company to focus their testing activities on identifying the vulnerability causing the anomaly.

The execution of some of the security controls shown in Table 8 cannot be automated (e.g., applying security updates and secure device configuration). For example, users can update specific vulnerable devices, or security engineers can modify the network configuration to prevent attacks in specific households. Our approach is the first step towards enduring security by supporting activities that allow us to detect and reason about unknown attacks.

A human-machine collaborative approach would instil a culture of continuous improvement and resilience against evolving cyber threats. In the context of homes and groups of homes, the engagement of stakeholders becomes imperative, as this can support enduring security. To support involvement of users and security/software engineers, there is a need to increase their situational awareness and design human-machine interactions supporting different levels of agency [70] depending on the stakeholder's role and expertise. This is a promising research direction that we aim to explore in future work requiring efforts from multi-discipline collaboration, such as cybersecurity, HCI, psychology, and AI [71].

### 5.4 Threats to Validity

**Internal Validity:** All the attacks considered in our study are network attacks. Consequently, the technique hasn't been evaluated against web and physical attacks. The effectiveness of the diagnosis technique is dependent on the contextual factors that can be identified from the system.

**External Validity:** While the iForest is a capable technique to model normal device behaviour and identify anomalies, the available datasets did not allow us to evaluate its effectiveness against attacks subtly different from benign network traffic. Our model of the smart home is general enough that it can be used with minor modifications to specific smart home contexts. However, similar partial models must be created for different use cases (e.g., public infrastructures). Further, we created our model manually, but we are currently exploring using inductive and neuro-symbolic learning to automate the creation of such models from network traces at runtime [13].

Even though our approach of system modelling, attack detection and diagnosis is generalisable to other application domains, all the steps described in this paper need to be repeated for the specific use case, as is often the case with cyber security applications. This would require a new threat and system model for the system under consideration, identification of suitable datasets that represent the attacks typically performed against the system, training a new machine learning model for the system (which could require other ML or deep learning approaches), and finally applying our diagnosis technique. As such, this exceeds the scope of work for our paper but is something we look forward to exploring in future work.

The performance of the abductive reasoning logic could be improved using parallel processing since the Clingo utility is run as a separate process with inputs from the anomaly detector. In future work, we also plan to implement a hierarchical search of the security requirements that should yield performance gains in the case of related requirements. The approach is resource-intensive since an iForest model is needed per device. While model inference is not as computationally demanding as training, this technique is better suited to running on a dedicated device in the network rather than a resource-constrained device such as a router.

## 6 CONCLUSION

Of the many challenges to providing sustainable security to a long-lived system such as a smart home, detecting unknown attacks as the system is used and diagnosing them in a manner that enables the enactment of suitable security controls is imperative. Unlike existing techniques focused on identifying new variants of known attacks or relying solely on anomaly detection, our approach prioritizes the system's security requirements as the foundation. By bridging the gap between anomalous behaviors indicating potential attacks and the desired security requirements, our technique shows promise in dealing with the vastly complicated category of unknown attacks.

In this paper, we proposed a three-step approach to address these challenges. Firstly, we modeled the smart home and formalised its security requirements using Answer Set Programming (ASP). Next, we developed an iForest-based anomaly detection mechanism to identify abnormal behaviors in the home network that could signify attacks. Finally, we implemented a diagnosis technique using abduction by refutation to diagnose the security requirements violated by network anomalies in the home. Evaluation of our technique using real-world datasets, including CICIoT2023 and IoT23, demonstrated its effectiveness in detecting & diagnosing threats. Our findings highlight the complex nature of security monitoring and decision-making tasks, emphasizing the importance of human intervention, particularly in attack detection and diagnosis.